\documentclass[sigconf, nonacm]{acmart}

\usepackage{algorithmic}
\usepackage{graphicx}
\usepackage{textcomp}
\usepackage{xspace}
\usepackage[usenames,dvipsnames]{xcolor}
\usepackage{tcolorbox}
\usepackage{makecell}
\usepackage{paralist}
\usepackage{listings}
\usepackage[utf8]{inputenc}
\usepackage{textcomp}
\usepackage{adjustbox}
\usepackage{colortbl}
\usepackage{multirow}

\newcommand{\lstbg}[3][0pt]{{\fboxsep#1\colorbox{#2}{\strut #3}}}
\lstdefinelanguage{diff}{
  basicstyle=\ttfamily\scriptsize,
  morecomment=[f][\lstbg{red!20}]-,
  morecomment=[f][\lstbg{green!20}]+,
  morecomment=[f][\textit]{@@},
}
\tcbuselibrary{skins}
\tcbset{
  tab3/.style=
  {
    enhanced,
    fonttitle=\bfseries,fontupper=\normalsize\sffamily,
    colback=white, 
    coltitle=white,
    center title,
    boxsep=0pt,
    left=0pt,
    right=0pt,
    top=0pt,
    bottom=0pt
  }
}

\newcommand{\eg}{\emph{e.g.,}\xspace}
\newcommand{\ie}{\emph{i.e.,}\xspace}

\newcommand{\etal}{\emph{et~al.}\xspace}
\newcommand{\secref}[1]{Section~\ref{#1}\xspace}

\newcommand{\figref}[1]{Figure~\ref{#1}\xspace}

\newcommand{\tabref}[1]{Table~\ref{#1}\xspace}

\newcommand{\gpt}{GPT-4o \emph{mini}\xspace}

\definecolor{lightlightgrey}{RGB}{233, 236, 239}

\AtBeginDocument{%
  \providecommand\BibTeX{{%
    Bib\TeX}}}

\newboolean{showcomments}
\setboolean{showcomments}{true}

\ifthenelse{\boolean{showcomments}}
  {\newcommand{\nb}[2]{
    \fbox{\bfseries\sffamily\scriptsize#1}
    {\sf\small$\blacktriangleright$\textit{#2}$\blacktriangleleft$}
   }
  }
  {\newcommand{\nb}[2]{}
  }

\newtcolorbox{promptbox}{colback=white, arc=0.5mm, top=1mm, bottom=1mm, left=1mm, right=1mm, title=System prompt used for code generation}

\newtcolorbox{optimizationbox}{colback=white, arc=0.5mm, top=1mm, bottom=1mm, left=1mm, right=1mm, title=Code optimization prompt}

\newtcolorbox{finalbox}{colback=white, arc=0.5mm, top=1mm, bottom=1mm, left=1mm, right=1mm, title=Prompt optimization}

\newtcolorbox{rq1box}{colback=white, arc=0.5mm, top=1mm, bottom=1mm, left=1mm, right=1mm, title=RQ$_1$ Summary}

\newtcolorbox{rq2box}{colback=white, arc=0.5mm, top=1mm, bottom=1mm, left=1mm, right=1mm, title=RQ$_2$ Summary}

\newtcolorbox{rq3box}{colback=white, arc=0.5mm, top=1mm, bottom=1mm, left=1mm, right=1mm, title=RQ$_3$ Summary}

\setcopyright{none} 
\renewcommand\footnotetextcopyrightpermission[1]{}

\def\BibTeX{{\rm B\kern-.05em{\sc i\kern-.025em b}\kern-.08em
    T\kern-.1667em\lower.7ex\hbox{E}\kern-.125emX}}
\begin{document}

\title{Guidelines to Prompt Large Language Models \\ for Code Generation: An Empirical Characterization}

\begin{abstract}
Large Language Models (LLMs) are extensively used nowadays for various software engineering tasks, primarily code generation. Previous research has shown how suitable prompt engineering could help developers improve their code generation prompts. However, so far, there are no specific guidelines driving developers to write suitable prompts for code generation.
In this work, we derive and evaluate development-specific prompt optimization guidelines.
We use an iterative, test-driven approach to automatically refine code generation prompts, and we analyze the outcome of this process to identify prompt improvement items that lead to test passes. We use such elements to elicit 10 guidelines for prompt improvement, related to better specifying I/O, pre/post conditions, providing examples, various types of details, or clarifying ambiguities. 
We conducted an assessment with 50 practitioners, who reported their usage of the elicited prompt improvement patterns, as well as their perceived usefulness.
Our results have implications not only for practitioners and educators, but also for those creating LLM-aided software development tools. 
\end{abstract}

\keywords{Large Language Models for Code Generation; Prompt Optimization}

\author{Alessandro Midolo}
\authornote{Both authors contributed equally to this research.}
\email{alessandro.midolo@unict.it}
\affiliation{%
  \institution{Dipartimento di Matematica e Informatica, University of Catania}
  \city{Catania}
  \country{Italy}
}

\author{Alessandro Giagnorio}
\authornotemark[1]
\email{alessandro.giagnorio@usi.ch}
\affiliation{%
  \institution{Software Institute – USI Università della Svizzera italiana}
  \city{Lugano}
  \country{Switzerland}
}

\author{Fiorella Zampetti}
\email{fzampetti@unisannio.it}
\affiliation{%
  \institution{University of Sannio}
  \city{Benevento}
  \country{Italy}
}

\author{Rosalia Tufano}
\email{rosalia.tufano@usi.ch}
\affiliation{%
  \institution{Software Institute – USI Università della Svizzera italiana}
  \city{Lugano}
  \country{Switzerland}
}

\author{Gabriele Bavota}
\email{gabriele.bavota@usi.ch}
\affiliation{%
  \institution{Software Institute – USI Università della Svizzera italiana}
  \city{Lugano}
  \country{Switzerland}
}

\author{Massimiliano Di Penta}
\email{dipenta@unisannio.it}
\affiliation{%
  \institution{University of Sannio}
  \city{Benevento}
  \country{Italy}
}

\maketitle
\section{Introduction}
The use of Large Language Models (LLMs) is radically changing how developers perform various software engineering tasks, including code generation, (re) documentation, software quality improvement, learning new pieces of technology, and others \cite{Tufano:msr2024, fakhoury2024llm, jki9ang2025survey, chen2024survey}.

Focusing in particular on code generation, on the one hand, the task seems to be pretty straightforward, \ie in several cases an LLM-based assistant is already able to generate a working solution based on a simple description of the task. 
At the same time, several studies \cite{Li:tosem2024,Khojah:tse2025} outlined how some prompt elements, for example, the system prompt and the role played by the LLM---\eg an expert Python or Java developer---may have a significant effect on the quality and correctness of the generated code. 

However, there may be several subtle elements that, if missed, may compromise the outcome of LLM-based code generation, such as the explanation of a method/function parameters, return values, pre- and post-conditions, or how exceptional cases are handled. To date, while general-purpose prompt engineering guidelines \cite{openai_playground_prompts} and automated prompt optimization approaches exist \cite{pryzant2023automatic,openai_playground_prompts}, there is a lack of specific guidelines and optimization tools targeting prompt engineering for software development. We conjecture that this is necessary because a software development prompt may need to contain very specific elements related to the technology, the specification, or the solution being requested, and this goes beyond general-purpose prompt optimization.

In this paper, we are paving the way toward providing developers with guidelines to improve their code generation prompts. To achieve this goal, we started with code generation tasks from three Python code generation benchmarks: BigCodeBench \cite{zhuo2024big}, HumanEval+ \cite{liu2023code}, and MBPP+ \cite{liu2023code}. We then created a simple prompt for four state-of-the-art LLMs---GPT-4o-mini \cite{gpt4omini}, Llama 3.3 70B Instruct \cite{llama3.3}, Qwen2.5 72B Instruct \cite{qwen2.5}, and DeepSeek Coder V2 Instruct \cite{deepseekv2}---and considered cases in which, over multiple iterations, the LLMs always generated code that led the benchmarks' test cases to fail. Then, we used an automated, iterative approach that results in the LLM coming up with a prompt that is able to generate a test-passing code.

After that, by manually analyzing the initial and final prompts and the test logs, we elicited a set of textual elements that the automated refinement added to the prompt to make the test cases pass. This allowed us to create a taxonomy of 10 code generation prompt improvement dimensions.

To evaluate the obtained catalogue of prompt optimization guidelines, we conducted a survey study involving 50 practitioners from our professional network. In the study, we asked participants about (i) the extent to which they use, in their development activities, the prompt optimization patterns we elicited, as well as other patterns we did not mention, and (ii) their perceived usefulness of these optimization patterns.  

Results of the study indicated varying usage of such patterns, \eg participants often tend to refine the I/O format or pre- and post-conditions, while they less often use a ``by example'' approach or perform linguistic improvements.  At the same time, they perceived particularly useful not only most of the optimization patterns they already use, but also some, such as those related to adding I/O examples, that they tend to use less so far.

The provided guidelines can be used, on the one hand, as a support for software practitioners and educators. On the other hand, in the future, they can lay the basis to create automated recommenders able to identify---also based on the context---missing elements from a prompt and suggest improvements for it.

\section{Related Work}

The growing complexity of LLM-based tasks has led to a surge in automated prompt optimization techniques. The recent survey by Ramnath \etal \cite{ramnath2025systematic} systematically categorizes such techniques—including search-based methods, reinforcement learning, and human-feedback loops—while identifying open challenges like alignment and generalization. In particular, Pryzant \etal \cite{pryzant2023automatic} introduce ProTeGi, which mimics gradient descent by generating natural-language feedback (textual gradients) to improve prompts iteratively. Wang \etal \cite{wang2023prompt} take a more autonomous approach with PromptAgent, an LLM-powered agent that formulates prompt optimization as a planning problem, combining evaluation, goal decomposition, and self-improvement. Jinyang \etal \cite{jinyang2025ddpt} propose DDPT, which employs a diffusion model to learn continuous prompt embeddings optimized for code quality.

While these techniques advance the automation of prompt refinement, they primarily operate as black boxes, offering limited transparency or actionable guidance to end users. In contrast, our work introduces a set of practical, empirically grounded guidelines, designed to support developers in crafting more effective prompts for code generation tasks, bridging the gap between sophisticated optimization methods and real-world LLM usage.

Several works emphasize the role of human insights in improving prompt quality. Liu \etal \cite{liu2023improving} demonstrate how iterative, chain-of-thought (CoT) prompt refinement can guide ChatGPT toward more accurate and context-aware outputs. Shin \etal \cite{shin2025prompt} compare prompt engineering to fine-tuning for code-related tasks and find that interactive, conversational prompting—rather than static prompts—leads to superior results. Similarly, Zhang \etal \cite{zhang2024instruct} propose an interactive prompting framework for code snippet adaptation, showing that model-driven questioning followed by refinement yields more reliable transformations. However, procedural techniques such as CoT inherently require longer prompt sequences, increasing token usage, computational cost, and inference latency. Moreover, extended interactions have been shown to exacerbate hallucination tendencies in LLMs~\cite{zheng2024navigate}. To address these limitations, our approach focuses on generating single, well-crafted prompts sufficiently detailed to elicit accurate and reliable responses—without relying on lengthy dialogue or iterative refinement.

Beyond procedural guidance, human feedback has also been employed in prompt optimization. Lin \etal \cite{lin2024prompt} introduce a preference-based optimization framework that refines prompts based on human comparisons, bypassing the need for task-specific labels. This aligns with Wen \etal \cite{wen2023hard}, who present a method for optimizing discrete (hard) prompts via gradient-based search, offering an interpretable and automatable alternative to manual crafting.

Understanding the properties of effective prompts is critical for both design and optimization. Mao \etal \cite{mao2025prompts} conduct a structured analysis of prompt templates, revealing how layout, placeholders, and component composition impact instruction-following behavior. Lee \etal \cite{lee2025predictive} introduce SPA, a syntactic analysis tool that predicts output characteristics from prompts without executing the LLM—enabling proactive prompt assessment.
Fagădău \etal \cite{fagadau2024analyzing} focus on prompt features in code generation, showing that structural cues like I/O examples and method summaries strongly correlate with higher-quality outputs. Likewise, Wu \etal \cite{wu2024chatgpt} analyze real-world ChatGPT usage patterns, linking prompt clarity and specificity with better developer outcomes. Our work complements this line of research, providing empirically-informed guidelines for prompt optimization in the context of code generation. 

Recent studies explore how prompts integrate into software development lifecycles. Liang \etal \cite{liang2025prompts} argue that prompts function as first-class programmatic artifacts, noting developers' challenges in debugging and understanding LLM behavior. Tafreshipour \etal~\cite{tafreshipour2025prompting} provide the first large-scale analysis of prompt evolution in codebases, uncovering common maintenance patterns and calling for better tooling and documentation. Complementing these artifact-level perspectives, Otten \etal \cite{otten2025prompting} empirically investigate how software developers employ generative AI tools in practice. Their survey of professional programmers reveals diverse prompting strategies across coding tasks—from generation to debugging and review—and highlights that iterative, conversational prompting dominates real-world use. These findings underscore a broader gap: inadequate documentation and guidance for effectively using LLMs in real-world inference tasks. Our work addresses this need by offering structured guidelines that help developers craft high-quality prompts, bridging the usability divide between LLM capabilities and practical software engineering applications.

In requirements engineering, Vogelsang \etal \cite{vogelsang2025impact} analyze how linguistic smells in requirements affect LLM-driven traceability tasks, finding that poor formulations can impair performance in nuanced ways. Abukhalaf \etal \cite{abukhalaf2024pathocl} tackle prompt scalability in model-to-text tasks with PathOCL, which selects relevant UML paths to keep prompts concise and accurate.

Several works highlight the importance of domain and context in crafting effective prompts. Peng \etal \cite{peng2024domain} propose TypeFix, which mines repair patterns to build domain-specific prompts for Python type error correction, outperforming rule- and learning-based baselines. Toufique \etal \cite{toufique2024automatic} enhance summarization by augmenting prompts with shallow semantic features, demonstrating that even lightweight program analysis improves LLM understanding.

Peng \etal \cite{peng2024reposim} introduce RepoSim, a benchmark for prompt strategies in code completion that captures realistic developer behavior. By leveraging contextual features like recent file edits and temporal patterns, RepoSim allows for more robust evaluation of prompt effectiveness under real-world conditions.

Pister \etal \cite{pister2024promptset} reinforce this view by presenting PromptSet, a dataset of developer-written prompts, and propose static analysis tools similar to linters to support prompt maintenance. Our work can further inform the design of these tools, thanks to the set of (validated) prompt-improvement guidelines we release.

\begin{figure*}[t]
    \centering
    \includegraphics[width=0.75\linewidth]{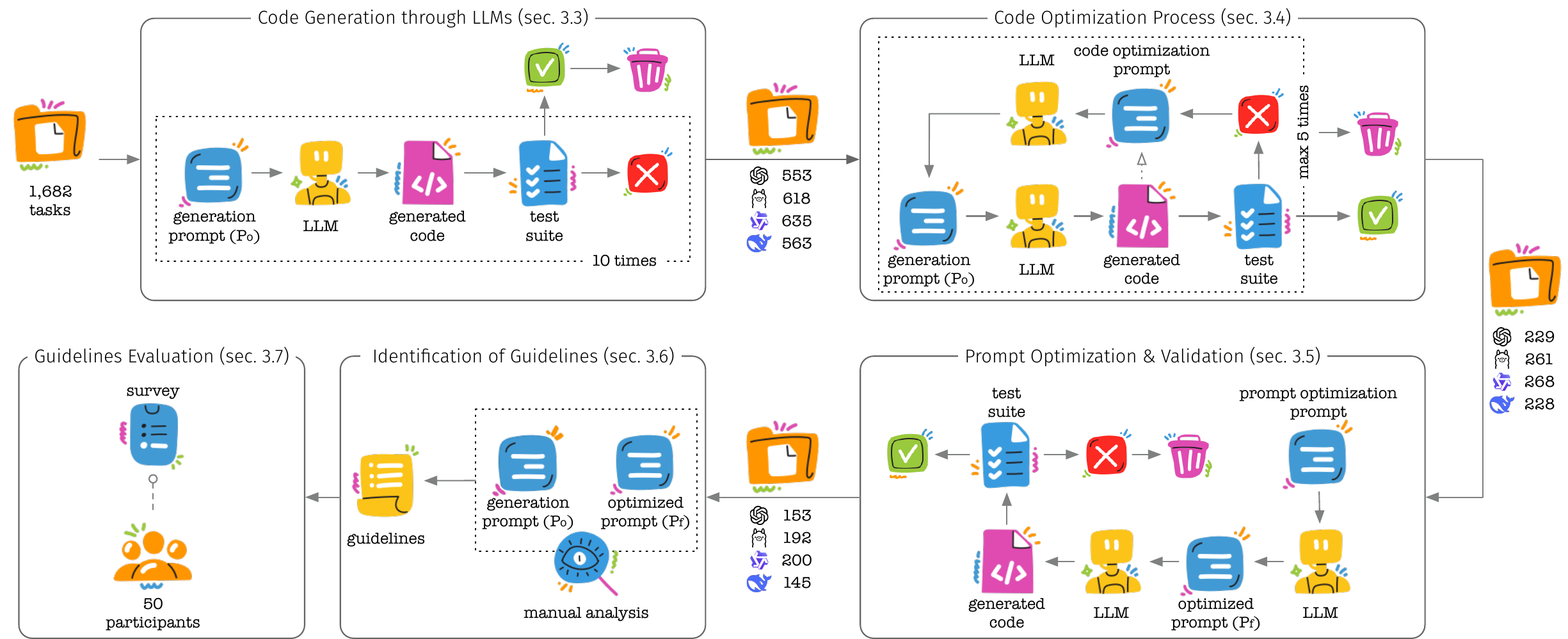}
    \caption{Methodology Overview}
    \label{fig:methodology}
    \vspace{-3mm}
\end{figure*}

\section{Study Design}

The \emph{goal} of this work is to establish guidelines for aiding developers in zero-shot LLM prompting in code generation. The \emph{perspective} is that of developers aiming to leverage natural language specifications for source code generation using LLMs. The \emph{context} consists of (i) development tasks from three benchmarks---BigCodeBench~\cite{zhuo2024big}, HumanEval+~\cite{liu2023code}, MBPP+~\cite{liu2023code}---and (ii) four state-of-the-art LLMs---\gpt \cite{gpt4omini}, Llama 3.3 70B Instruct \cite{llama3.3}, Qwen2.5 72B Instruct \cite{qwen2.5}, DeepSeek Coder V2 Instruct \cite{deepseekv2}.

\newcommand{\rqone}{What is the set of information to be included in the prompt to maximize the chance of success in a code generation task?}
\newcommand{\rqtwo}{To what extent do the provided guidelines reflect the prompt improvement strategies used by practitioners?}
\newcommand{\rqthree}{To what extent do practitioners perceive the provided prompt improvement guidelines as useful?}

The study aims to address the following research questions:
\begin{itemize}
\item {\bf RQ$_1$:} \emph{\rqone} RQ$_1$ aims at identifying---through an automated prompt optimization process and a follow-up manual analysis---elements that might be necessary to add in a code generation task prompt to generate a correct code.
\item {\bf RQ$_2$:} \emph{\rqtwo} With this research question, we investigate the extent to which study participants have used, in their development activities, the prompt improvement guidelines identified in RQ$_1$.
\item {\bf RQ$_3$:} \emph{\rqthree} In RQ$_3$, we assess the study participants' perceived usefulness of the provided guidelines, regardless of whether or not they use them.
\end{itemize}

\subsection{Benchmark description}
For our study, we leverage tasks from three benchmarks created for evaluating code generation models in Python: BigCodeBench~\cite{zhuo2024big}, HumanEval+~\cite{liu2023code}, and MBPP+~\cite{liu2023code}. BigCodeBench assesses LLMs on real-world programming tasks, focusing on complex reasoning and multi-library code generation. It comprises 1,140 Python tasks covering 139 libraries across seven domains, requiring models to compose solutions using diverse function calls. Each task includes an average of 5.6 test cases to ensure rigorous evaluation. HumanEval+ and MBPP+ extend HumanEval \cite{humaneval} and MBPP \cite{mbpp} respectively, by significantly increasing the number of test cases—80$\times$ more for HumanEval+ and 35$\times$ more for MBPP+. They feature 164 (HumanEval+) and 378 (MBPP+) coding tasks, evaluating functional correctness in LLM-generated code.

\subsection{Methodology overview}
\label{sec:methodology}
\figref{fig:methodology} provides an overview of our methodology. For each benchmark's task and each LLM independently, we ask the LLM ten times to generate a valid code implementation. We then select the coding tasks for which the LLM always failed to generate a correct implementation (according to test execution) and start an optimization process guiding the LLM to generate the correct code based on the test results. If, within a maximum number of iterations, the LLM manages to output a correct implementation, this leads to a triplet <$P_o$, [$C_w$], $C_c$>, where: (i) $P_o$ is the original code generation prompt provided in the benchmark; [$C_w$] is a set of wrong implementations outputted by the LLM when prompted with $P_o$ during the run of the iterations, with each (wrong) implementation paired with the errors printed by the tests; and (iii) $C_c$ is the fixed version of the code that the LLM generated when we showed it the errors reported by the tests. Such a triplet is finally provided to the LLM, asking it to automatically revise $P_o$ to produce a ``fixed'' prompt ($P_f$)  which is likely to generate $C_c$ (\ie the correct implementation). The generated $P_f$ is then prompted to the LLM to verify whether it actually results in a correct implementation. If this is the case, such a process provides us with pairs <$P_o$, $P_f$> (\ie original and fixed prompt) which we manually inspect, coming up with a catalogue of prompt optimizations implemented by the LLMs. Such a catalogue has been finally validated in a human study with practitioners.

Note that the whole procedure we use to generate $P_f$ is not meant to be an automated process used by developers in practice, since it assumes the availability of test execution results, which are often not available during code generation. The procedure is meant to automatically build a dataset of prompt optimizations to be manually inspected in our study. 

In the following, we detail the main steps of our methodology.

\subsection{Code Generation through LLMs}
\label{sub:generation}
The initial phase of our study focuses on source code generation using four state-of-the-art LLMs: \gpt \cite{gpt4omini}, Llama 3.3 70B Instruct \cite{llama3.3}, Qwen2.5 72B Instruct \cite{qwen2.5}, and DeepSeek Coder V2 Instruct \cite{deepseekv2}. We selected these models because recent studies have demonstrated their effectiveness in various software engineering applications \cite{ehsani2025bug, fan2025sek, zhang-etal-2025-building, golubev2025training}. While other, even more performant models could be used, these models proved to be suitable for our purpose, \ie to iteratively optimize code to make test passing, finally leading to an optimized prompt. As input to the LLMs, we adopt the prompts provided in the subject datasets (\ie BigCodeBench, HumanEval+, and MBPP+), which consist of the method signature and its docstring. The signature is crucial to ensure that the generated method can be executed within the test suite without issues related to the method or parameter names.

\textbf{Generating the Code}.  Following recent studies~\cite{dong2024self, xueying2024evaluating}, we performed the generation process with zero temperature and a fixed seed. Lowering the temperature reduces nondeterminism compared to the default configuration~\cite{ouyang2025empirical}, and has been shown to improve performance in code generation tasks~\cite{arora2024optimizing, doderlein2022piloting}. However, since the models do not guarantee full determinism even at low temperatures, we generated ten outputs for each LLM and for each task. Each generated code has then been validated against the related test suite. All the coding tasks for which each LLM always generated test-failing solutions are kept for the subsequent steps of our methodology. \tabref{tab:generation} shows the breakdown of failing coding tasks per LLM and benchmark. Note that we had to exclude one task from HumanEval+ and three tasks from MBPP+ since their tests were not providing error messages when failing, thus not allowing their usage in the optimization process, resulting in 163 coding tasks for HumanEval+ and 375 for MBPP+.

\begin{table}[t!]
    \caption{Code generation results with ten runs per task: Number of tasks that passed at least once and always failed. The total number of tasks is  1,140 for BigCodeBench, 163 for HumanEval+, and 375 for MBPP+\vspace{-0.4cm}}
    \label{tab:generation}
    \small
    \begin{tcolorbox}[tab3, title={\vspace{0.05cm}Code Generation Results\vspace{0.05cm}}, boxrule=0.5pt, width=0.95\linewidth]
    \begin{tabular*}{\linewidth}{@{\hspace{6pt}}l@{\extracolsep{\fill}}rr@{\hspace{6pt}}}

    \rowcolor{lightlightgrey}
    \multicolumn{3}{c}{\textbf{\small{GPT-4o mini}}} \\
    \toprule
    \textbf{Benchmark} & \textbf{\#At least one pass} & \textbf{\#Always fail} \\
    \midrule
    HumanEval+    & 145 & 18 \\
    MBPP+         & 265 & 110 \\
    BigCodeBench  & 715 & 425  \\
    \bottomrule

    \rowcolor{lightlightgrey}
    \multicolumn{3}{c}{\textbf{\small{Llama 3.3 70B Instruct}}} \\
    \toprule
    \textbf{Benchmark} & \textbf{\#At least one pass} & \textbf{\#Always fail} \\
    \midrule
    HumanEval+    & 132 & 31 \\
    MBPP+         & 248 & 127 \\
    BigCodeBench  & 680 & 460  \\
    \bottomrule

    \rowcolor{lightlightgrey}
    \multicolumn{3}{c}{\textbf{\small{Qwen2.5 72B Instruct}}} \\
    \toprule
    \textbf{Benchmark} & \textbf{\#At least one pass} & \textbf{\#Always fail} \\
    \midrule
    HumanEval+    & 134 & 29 \\
    MBPP+         & 258 & 117 \\
    BigCodeBench  & 651 & 489  \\
    \bottomrule

    \rowcolor{lightlightgrey}
    \multicolumn{3}{c}{\textbf{\small{DeepSeek Coder V2 Instruct}}} \\
    \toprule
    \textbf{Benchmark} & \textbf{\#At least one pass} & \textbf{\#Always fail} \\
    \midrule
    HumanEval+    & 134 & 29 \\
    MBPP+         & 276 & 99 \\
    BigCodeBench  & 705 & 435  \\

    \end{tabular*}
    \end{tcolorbox}
    \vspace{-0.25cm}
\end{table}

\subsection{Code Optimization Process}

We perform an iterative code optimization inspired by the approach by Pryzant \etal \cite{pryzant2023automatic}. We utilize the LLMs to generate new code based on feedback derived from test failures, until the code passes the tests or a maximum number of optimization attempts have been made. The optimization process begins with the code generation prompt we described in \secref{sub:generation}, which is again provided to the LLMs to get the output code. Remember that this is done only for the coding tasks for which the same prompt provided to the same LLM already failed ten times in generating a correct implementation (as described in \secref{sub:generation}). Thus, we expect that the specific LLM would fail this time as well. Still, we assess the code against the corresponding test suite. If (surprisingly) the tests pass, the optimization process stops, and this instance is discarded (this never happened in our experimentation.) Otherwise, we provide the LLM the following code-optimization prompt:

\begin{optimizationbox}
I'm trying to write a zero-shot prompt for code generation.
This is the [round] round of optimization.
My starting prompt is:
"[prompt]"
In the last round of optimization, you provided this code snippet:
"[code]"
But this source code does not pass the test cases, failing with the following messages:
"[error]"
Based on the above information, provide a new code snippet that satisfies the requirements and passes the test cases.
\end{optimizationbox}

The code-optimization prompt contains four tags: ``round'' indicates the $i^{th}$ iteration of the optimization process; ``prompt''refers to the starting code generation prompt; ``code'' is the code generated in the previous round of optimization (or the code produced from the initial prompt in the first round); and ``error'' contains the error messages from testing the previously generated code. The code outputted by the LLMs is again tested and, in case of failure, another round of code optimization is triggered. If the LLM is unable to generate a valid source code within five rounds, the process halts, and the coding task is discarded. In case of code-passing tests, instead, we store: (i) the original prompt used for the code generation; (ii) all test error messages collected during the code optimization rounds; (iii) the test-failing implementations seen during the process; and (iv) the test-passing code which stopped the process. This process is performed for each LLM independently, so it is possible that some tasks are discarded for one LLM but kept for another. 

A total of 229, 261, 268, and 228 instances were successfully optimized by GPT-4o mini, Llama 3.3, Qwen2.5, and DeepSeek, respectively, thus progressing to the next step of prompt optimization.

\subsection{Prompt Optimization \& Validation}
At this point, we instruct each LLM to provide a new zero-shot prompt by learning from the information derived from the code-optimization process. We prompt the LLMs as follows:

\begin{finalbox}
I'm trying to write a zero-shot prompt for code generation.
This is the [round], and last, round of optimization.
My starting prompt is:
"[prompt]"
In the previous steps of optimizations, the codes you provided did not pass the test cases, failing with the following messages:
[error]
In the last rounds of optimization, you provided this code snippet:
"[code]"
This code successfully passed all test cases.
Based on the above information, provide a new prompt that, differing from the starting prompt, will provide enough details to generate a code snippet that passes the test cases.
Wrap the new prompt between triple quotes.
\end{finalbox}

This prompt includes four customizable tags:  ``round'' indicates the $i^{th}$ and last iteration of the optimization process; ``prompt'' is again the starting prompt used in the code generation process; ``error'' is a list of pairs <code, errors>, where each pair corresponds to a test-failing code generated during the code optimization process, reporting the generated code and the corresponding test failures. Finally, ``code'' is the test-passing code output of the code optimization process. This provides the LLMs with contextual information related to all previously generated code, offering a comprehensive view of the information missing in the original prompt.

The fixed prompts $P_f$ outputted by the LLMs, which should result in test-passing codes, are provided ten times as input to the LLMs to verify whether the resulting codes are actually correct. If any of those ten runs result in passing tests, we consider $P_f$ as a successfully optimized prompt, and pair it with the original prompt $P_o$ (\ie the one used in the very first code generation step --- \secref{sub:generation}). Such pairs constitute the input for the subsequent manual inspection aimed at identifying prompt guidelines. In total, GPT-4o mini, Llama 3.3, Qwen2.5, and DeepSeek succeeded in creating an optimized prompt for 153, 192, 200, and 145 instances, respectively.

\begin{figure}[t]
\centering
\noindent
\begin{minipage}[c]{0.48\textwidth} 
\begin{adjustbox}{valign=c}
\begin{lstlisting}[caption={Original prompt}, label={}]
def decimal_to_binary(n):
    """Write a function to convert the given decimal number to its binary equivalent, represented as a string with no leading zeros."""
\end{lstlisting}
\end{adjustbox}
\end{minipage}%

\begin{minipage}[c]{0.48\textwidth}
\begin{lstlisting}[caption={Optimized prompt}, label={}]
def decimal_to_binary(n):
    """Convert a given decimal integer to its binary representation as a string. 
    The function should handle both positive and negative integers, returning 
    the binary equivalent without leading zeros. For negative integers, 
    the binary representation should be prefixed with a '-' sign. 
    If the input is zero, return '0'."""

\end{lstlisting}
\end{minipage}
\vspace{-0.35cm}
\caption{Original (top) and optimized (bottom) prompt. Example from the mbpp dataset, task id: 99, LLM used: \gpt \vspace{-0.5cm}}
\label{fig:prompt_comparison}
\end{figure}

\figref{fig:prompt_comparison} illustrates a representative example of prompt refinement. The original prompt provides only a general task description, which leaves several ambiguities in the expected behavior of the generated code. In contrast, the optimized prompt explicitly specifies \textit{input/output formats} and \textit{post-conditions}, defining precise behavior for positive, negative, and zero inputs. These clarifications ensure that the returned value satisfies well-defined constraints, improving both predictability and correctness. 

\subsection{Identification of Guidelines} \label{sec:guidelines}
We manually inspect all instances for which LLMs were able to generate a test-passing prompt, starting from a test-failing one. 

We divided all optimized prompts into three groups, each assigned to two authors acting as inspectors. Within each group, the set of prompts was split between the two inspectors: each analyzed their own subset and a counter-analysis of the other inspector's subset. This makes every classification double-checked. 

To perform the analysis, each inspector had access to:
\begin{enumerate}
\item A diff file (in HTML) highlighting the differences between each starting prompt and its optimized version.
\item The test log, reporting the failures that occurred during the iterative process.
\end{enumerate}

In total, 224/627 of the inspected instances resulted in a conflict (\ie a different set of optimization guidelines defined by the two inspectors). Such a high number of conflicts is expected since we did not start from a pre-defined catalogue of guidelines, but we derived it as we moved on in the process. Afterwards, each pair of inspectors resolved the conflicts in their classifications through open discussion. To define the guidelines, the evaluators identified the elements of the prompt that were added, removed, or changed. At the same time, they examined whether the changes (i) included the implementation of the code itself, or (ii) incorporated too specific requirements which could only be known by reading the tests.
We excluded the first category, as the inclusion of implemented code could introduce bias into the generation of new prompts during the optimization phase. We also excluded the second since those contained pieces of information not available in the original task specification, which a developer could only have access to in a test-driven development scenario.

Once the manual analysis was completed, two authors scrutinized the list of all possible guidelines identified and consolidated it by merging very similar ones where appropriate. In the end, we obtained a set of 10 guidelines.

\subsection{Guideline Evaluation}
To evaluate the elicited guidelines, we conduct a study in which we ask practitioners the extent to which they (i) have used, in their code generation tasks, the prompt improvement strategies from our guidelines, as well as other optimizations we may have missed, and (ii) perceive each guideline as useful, regardless of their actual usage. The study participants have been recruited through the authors' professional network.

\begin{table*}[t!]
	\centering
	\small
	\caption{Guidelines proposed to improve prompts during code generation\vspace{-0.35cm}}
	\label{tab:guidelines}
     \begin{tcolorbox}[tab3, title={\vspace{0.05cm}Guidelines\vspace{0.05cm}}, boxrule=0.5pt, width=0.921\linewidth]
	\begin{tabular}{lp{11.5cm}r}
		\toprule
		\textbf{Guideline} & \textbf{Description} & \textbf{Applied}\\
		\bottomrule
		\cellcolor{lightlightgrey}\multirow{2}{*}{Requirements} & \cellcolor{lightlightgrey}Make requirements explicit in terms of needed packages or libraries, and explain what to use them for. & \cellcolor{lightlightgrey}19\% \\[1pt]
		
		Pre-conditions & Specify pre-conditions (\eg a data structure provided as input must be non-empty). & 7\% \\[1pt]
		
        \cellcolor{lightlightgrey}Post-conditions & \cellcolor{lightlightgrey}Specify post-conditions (\eg the return value is expected to be in a certain range). & \cellcolor{lightlightgrey}23\% \\[1pt]
        
        I/O format & Specify input/output and return format (if complex), and document special cases. & 44\% \\[1pt]
        
        \cellcolor{lightlightgrey}\multirow{2}{*}{Exceptions} & \cellcolor{lightlightgrey}If exceptions must be raised or error messages must be printed, be specific in reporting which exceptions to raise and how, or which error messages to write and where. & \cellcolor{lightlightgrey}12\% \\[1pt]
        
        \multirow{2}{*}{Algorithmic Details} & Detail algorithmic aspects of the code, ensuring the correctness of complex features with supporting notes if necessary, and include definitions or formulas for clarity and accuracy. & 57\% \\[1pt]
        
        \cellcolor{lightlightgrey}\multirow{2}{*}{Variable Mentioning} &\cellcolor{lightlightgrey} Don't use generic terms to refer to different variables or different terms to the same variable. &\cellcolor{lightlightgrey} 3\% \\[1pt]
        
        \multirow{2}{*}{Unclear Conditions} & When specifying a condition, do not use ``otherwise'' referring to the second condition. Instead, describe both conditions explicitly. & 1\% \\[1pt]
        
        \cellcolor{lightlightgrey}More Examples & \cellcolor{lightlightgrey}Add some examples of run (\ie doctests). &\cellcolor{lightlightgrey} 24\% \\[1pt]
        
        \multirow{2}{*}{Assertive Language} & For required functionality, use non-ambiguous language, \eg use ``must'' instead of ``should''. Use more assertive language. & 9\% \\[1pt]

	\end{tabular}
    \end{tcolorbox}
    \vspace{-2mm}
\end{table*}

The study has been administered as a survey questionnaire (available in our replication package \cite{replication}) conducted using Google Forms. The questionnaire structure features three sections in which we collect:

\begin{compactenum}
\item \emph{Background information:} This includes study title, development experience in years, and experience with the use of LLMs in software development;
\item \emph{Usage of the studied prompt improvement strategies:} we use a 5-level scale in which we ask the participants to estimate the frequency (\ie \% of code generation prompts they write) in which they use the considered guideline (Never, $<$25\%, between 25\% and 50\%, between 50\% and 75\%, over 75\% of the prompts). Also, we ask them to specify whether they use any other prompt improvement strategy that is not considered in our guidelines.
\item \emph{Perceived usefulness of the studied prompt improvement strategies:} In this case, we ask---using a 5-level Likert scale \cite{oppenheim1992questionnaire} from ``completely useless'' to ``very useful''---the \emph{perceived} level of usefulness of each prompt improvement guideline, regardless of its usage.
\end{compactenum}

We kept the questionnaire open for three weeks. Out of about 70 invitations, we collected 50 valid responses.

\textbf{Questionnaire demongraphics.} The study participants are pretty diversified in terms of expertise. This may reflect a diverse population of developers who may interact with LLMs to generate code. Nowadays, even quite inexperienced programmers do that.

Most of the participants held an M.Sc. (19) or a B.Sc. (19), whereas 9 have a Ph.D. and 3 are students with a high school degree. The development experience is between 1 and 5 years (22) or between 5 and 10 (20), with 3 participants with over 10 years of experience and 5 with less than one year of experience. Most of them use ChatGPT (45), followed by GitHub Copilot (24), Gemini (14), Claude (10), DeepSeek (4), Perplexity (2), and JetBrains AI assistant (1).

Participants report using LLMs in less than 25\% of their development activities (11), 25-50\% (22), 50-75\% (8), and over 75\% (9). 

\section{Results Discussion}
In the following, we discuss results addressing the three RQs.

\subsection{RQ$_1$: Identified Guidelines}
\label{sub:results_rq1}

\tabref{tab:guidelines} overviews the 10 guidelines we derived. The table indicates name, short description, and the percentage of instances (out of the 627 for which the LLMs were able to generate an optimized prompt) in which the LLMs applied such an optimization pattern. 
We detail the guidelines in the following. The ``ID" allows for tracing the examples through our replication package \cite{replication}.

\textbf{Requirements}. Explicitly state all requirements needed for the code to function correctly. It is also important to explain the purpose of each dependency (\eg ``use \texttt{numpy} for numerical array operations'' rather than simply ``import numpy''). This makes it more likely that the generated code includes correct imports and uses each dependency appropriately. Such a guideline has been applied in 19\% of optimized prompts. A concrete example of prompt optimization made by the LLMs by adopting this guideline concerns an instance from the BigCodeBench benchmark (ID: 125), in which the original prompt only listed a set of requirements to be used (\ie \texttt{Requirements: collections.defaultdict}) while the optimized one by \gpt explained the role played by each dependency (\ie \texttt{Use collections.defaultdict to count letter occurrences}).\smallskip

\textbf{Pre-conditions}. Define conditions that must hold before execution.  For example, specify that ``the input list must be non-empty'' or ``the matrix must be square.''  This helps avoid invalid assumptions and ensures robust error handling. Such a guideline, applied in 7\% of optimized prompts, has been adopted by GPT-40 mini when implementing a function to backup a given source folder to the specified backup directory. The LLM added to the prompt ``\texttt{Ensure that the backup directory is writable and has sufficient space}'', to avoid writing exceptions observed in previous, unsuccessful, implementations.\smallskip

\textbf{Post-conditions}. Specify a condition that shall be valid after code execution, \eg guarantees about the produced outputs (\eg ``the function returns a sorted list in ascending order'' or ``the result must be within [0, 1]''). Explicit post-conditions help the model verify the correctness of its output logic. The LLMs applied such a guideline in 23\% of the optimized propts. 

As an example, specifying a post-condition was needed for one of the MBPP+ tasks (ID 99), asking for the conversion of a decimal number into its binary format. The original prompt does not specify that "\texttt{for negative integers, the binary representation should be prefixed with a `-' sign}". Such a sentence has been added by \gpt in the optimized prompt.\smallskip

\textbf{Input/Output Format}. Clarify input and output data types, shapes, and structures, including any special or edge cases. The application of such a guideline is crucial, especially when dealing with more complex I/O structures. This is the second-most applied guideline, with 44\% of optimized prompts involved. For example, one of the always-failing coding tasks from BigCodeBench (ID 267) asked the LLM to plot specific information from a data dictionary. However, there were missing specifications about the expected format of such a plot. The following additions resulted in a test-passing code: \texttt{The plot should have: - X-axis labeled as `Frequency [Hz]'; - Y-axis labeled as `Frequency Spectrum Magnitude'}.\smallskip

\textbf{Exceptions and Error Handling}. If the code must handle errors or raise exceptions, describe these explicitly. Specify which exceptions to raise and under what conditions. If error messages must be printed, describe what message to print and where. Avoid vague wording like ``handle the error gracefully''; instead, define explicit exception types and messages. The specification of possible errors could relate to several parts of the coding tasks to be accomplished, and has been employed in 12\% of the optimized prompts. We observed \gpt optimizing the prompt (ID 560) from bigcodebench to include cases related to unexpected input parameters (\eg \texttt{This function plots a bar chart of monthly data values for a single year, with `month' on the x-axis and `value' on the y-axis. The function should raise a ValueError if the data contains entries from multiple years.}), as well as to unrespected precoditions in bigcodebench (ID 150),\eg \texttt{Ensure that if a product key in product\_keys does not exist in product\_dict, a KeyError is raised.}. In some cases, the exceptions to raise were already documented in the original prompt, but not \emph{all} scenarios causing the raising of the exception were documented (\eg \gpt added ``\texttt{or if the script file does not exist}'' to a sentence explaining cases in which a \texttt{ValueError} should be reported, bigcodebench, ID 460). This case shows how minor specifications in a prompt may make the difference between a test-passing and a test-failing implementation.\smallskip

\textbf{Algorithmic details}. Provide essential algorithmic details for complex logic. It is important to include supporting definitions, formulas, or theoretical notes where needed to ensure correctness. This helps the generated code follow the intended method rather than guessing or simplifying, and it is the most frequently applied prompt optimization guideline (57\% of cases). In one prompt optimized using this guideline (MBPP+, ID 781), \gpt expanded the original prompt (\ie \texttt{Write a python function to check whether the count of divisors is even}) by adding: \texttt{The function should efficiently count divisors by iterating only up to the square root of n}. In another case, \gpt clarified the definition of Jacobsthal number, which was not present in the original prompt, by adding: \texttt{The Jacobsthal numbers are defined as follows: J(0) = 0, J(1) = 1, and for n > 1, J(n) = J(n-1) + 2 * J(n-2)} (MBPP+, ID 752).\smallskip

\textbf{Variable mentioning}. Maintain consistent terminology. Do not use generic or interchangeable terms for different variables. Use the same term consistently for the same variable throughout the prompt. This minimizes confusion and ensures variable names and references remain coherent. While only 3\% of the optimized prompts benefited from such a guideline, it is interesting to see how simple linguistic variations helped code generation. In one of the original prompts of BigCodeBench (ID 731), the function to be generated takes as input two parameters named \texttt{data} and \texttt{target}. However, in the description of the code to implement, the original prompt refers to these two parameters as ``data'' (coherent with the parameter's name) and ``destination'' (not coherent) --- \texttt{Save the Sklearn dataset ("Data" and "Destination") in the pickle file}. \gpt revises this inconsistency by rephrasing the sentence as: \texttt{Save the provided Sklearn dataset (data and target) into a pickle file}.\smallskip

\textbf{Clarity in conditions}. Avoid ambiguous phrasing such as ``otherwise'' when describing conditional logic.  Instead, specify both conditions explicitly (\eg ``If \texttt{x > 0}, do A; if \texttt{x <= 0}, do B'').  This prevents the model from making incorrect assumptions about conditional relationships. Typical fixes of the LLMs included the replacement of the ``otherwise'' when describing the second condition of a selection construct, \eg from BigCodeBench (ID 477) optimized by \gpt: 

\texttt{[...] if N is greater than or equal to the number of categories, otherwise it is randomly sampled without replacement from CATEGORIES}

to:

\texttt{[...] if N is greater than or equal to the number of categories. If N is less than the number of categories, sample without replacement.} 

We only found a few cases of such an optimization (1\% of the instances).

\smallskip

\textbf{More examples}. Provide concrete examples or doctests demonstrating expected behavior. Examples make the functionality clearer and serve as an informal test of correctness. The examples added by the LLMs in 24\% of the optimized prompts often aim at documenting possible corner cases that must be handled. For example, in a function aimed at building a pandas DataFrame based on two lists provided as input, \gpt added the last three of the four usage examples shown below, with the first one already part of the original prompt, and task\_func being the name of the function to implement (BigCodeBench, ID 553):

\texttt{task\_func([1, 2, 3], [`A', `B', `C', `D', `E'])}

\texttt{task\_func([], [])}

\texttt{task\_func([1, 2, 3], [])}

\texttt{task\_func([], [`A', `B', `C'])}\smallskip

\begin{figure*}[t!]
    \centering
    \includegraphics[width=0.8\linewidth]{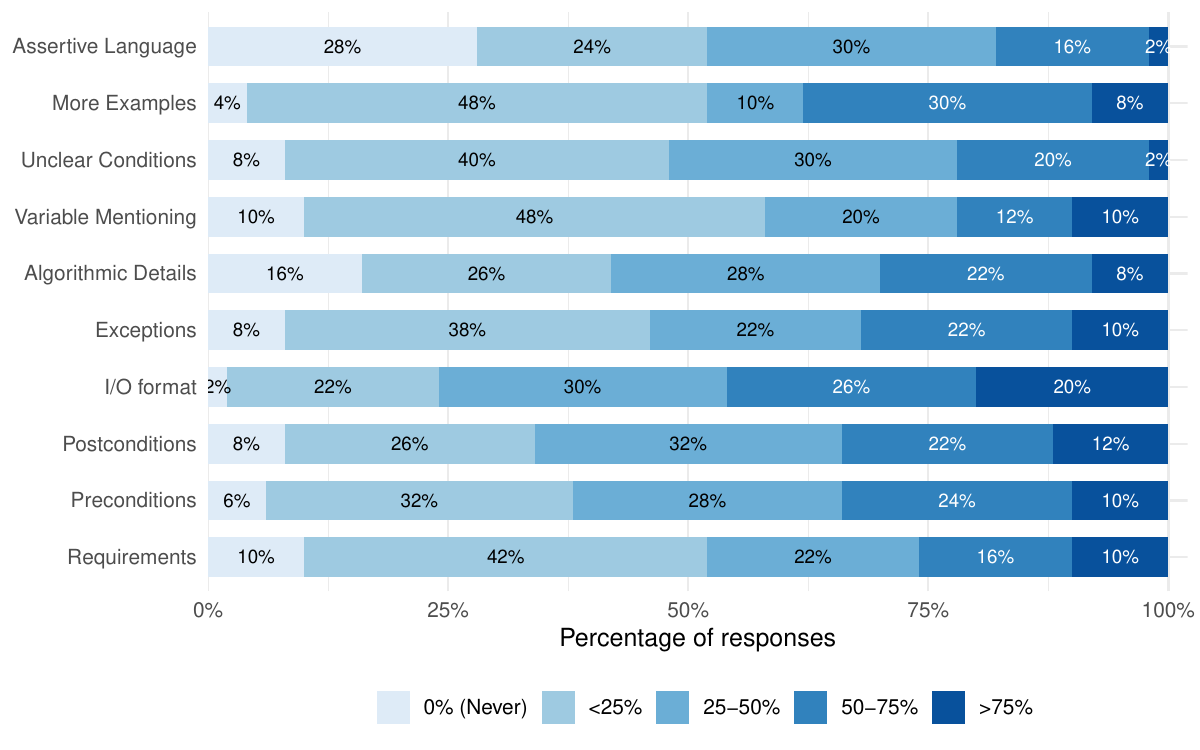}
    \caption{Prompt improvement patterns usage frequency}
    \label{fig:patternUsage}
\end{figure*}

\textbf{Assertive Language}. Use precise, assertive language to express requirements and constraints.  Prefer ``must'' or ``is required to'' instead of ``should'' or ``may.''  This reduces uncertainty and may increase the chances that the generated code adheres strictly to the specification. Such a guideline has been implemented by the LLMs in 9\% of optimized prompts, by injecting assertive terms such as ``ensure that''. \smallskip

\begin{rq1box}
 We elicited 10 guidelines for code generation prompt optimization. These were related to aspects concerning key aspects of code behavior, such as I/O format, and pre/post conditions, but also detailed and technological aspects, such as requirements, unclear conditions, and exceptions, and, finally, textual ambiguities.
\end{rq1box}

\subsection{RQ$_2$: Improvement pattern usage by the study participants}

When we asked participants how frequently they perform prompt improvements, they reported doing it in 25\% of their prompts (11), 25-50\% (21), 50-75\% (12), and over 75\% (6).

\figref{fig:patternUsage} shows a plot depicting the extent to which the study participants used the different prompt improvement patterns emerging from our guidelines. Results are reported with stacked bars with different color intensity representing the usage percentage of each prompt improvement pattern. In interpreting them, one should not expect a pattern to be used \emph{in the majority} of the prompts, as it may only apply to specific scenarios. 

\emph{Requirements}, \ie library imports, have a relatively mild use. Only 10\% report specifying them in over 75\% of the prompts, 16\% in 50-75\%, and 22\% in 25-50\%  of them. This may be since our respondents do not interact with LLMs having a specific technology in mind, unless they are already using some and need to enforce it. Therefore, they expect the LLM to provide the requirements itself. 

Both \emph{pre-conditions} and \emph{post-conditions} are among the patterns with a relatively higher frequency of use, with the majority of respondents indicating that they use them at least in 25-50\% of the prompts, even though the very frequent use (over 75\% of the prompts) is limited to 10\% of the respondents.  This is expected when one wants to generate a non-trivial method/function and wants to check the inputs or the produced outputs.

Even more used is the specification of the \emph{I/O format}. Especially (and not only) for data processing functions, this allows an easier integration of the generated code with the rest of one's own system, or in general, an easier usage of the function. 

Less used are patterns aimed at specifying \emph{exceptions}. While the majority uses it in at least 25-50\% of the prompts, only 10\% do so in over 75\% of them. This may depend either on the need to adapt exception handling to the rest of the system or on the fact that during code generation, handling exceptional cases is not the priority, except for specific functions. 

Similar results are observed for \emph{algorithmic details}, which may apply only to functions whose behavior needs to be explicitly specified, and cannot be inferred from the general description. 

Improving prompts to maintain a consistent terminology for the used \emph{variables} is done by only 10\% of the participants on over 75\% of the prompts, 12\% in 50-75\% of the prompts, and 20\% in 25-50\% of them. This is more of a linguistic optimization, which resulted from our analysis, yet respondents did not think about it in their activities. This may also be because the starting prompts used by participants feature explicit variable names, but more of the code logic.

\begin{figure*}[t!]
    \centering
    \includegraphics[width=0.8\linewidth]{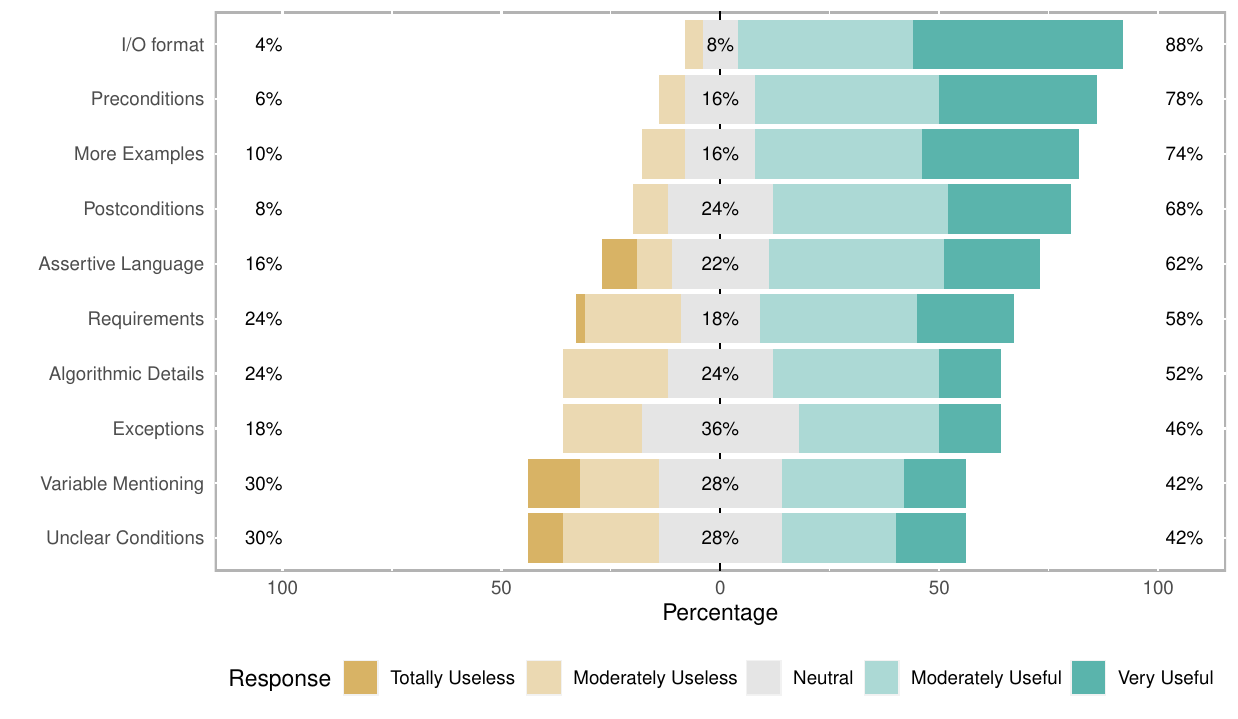}
    \caption{Perceived usefulness of the pattern improvement guidelines.}
    \label{fig:usefulness}
\end{figure*}

\emph{Unclear conditions} tend to be improved by the majority of respondents in at least 25-50\% of the cases, although only 2\%  report doing so for over 75\% of the prompts. Such needs for improvement are a circumstance that may occur only in specific code generation scenarios/functions where conditions are particularly complex or ambiguous. 

Surprisingly, specifying \emph{more examples} was reported to be used in 25-50\% of the cases by less than the majority of respondents (overall, 48\% of them), with only 8\% doing so in over 75\% of the prompts. This may even depend, once again, on the specific development scenario, but also on the limited attitude in doing so and in leveraging examples when writing code, \eg in a context of test-driven or behavior-driven development~\cite{karpurapu2024bdd, mathews2024tdd, liang2025exploring}.

The least used pattern relates to using \emph{assertive language}. This confirms, in a more general meaning, what found about \emph{variable} mentioning, \ie a limited use of linguistic-related improvements.

Besides reporting usage frequencies for patterns related to our guidelines, we also asked participants to report other prompt improvement strategies not present in our guidelines. They reported the following:
\begin{compactitem}
\item More contextual information (seven responses), including signatures of other methods/functions in the system/file or information about other components. Clearly, this also depends on the tool being used, \eg some LLM-based assistants such as GitHub Copilot are better capable of capturing context than Web-based ones;
\item Coding standards (two responses);
\item Testability requirements (one response); and
\item Non-functional aspects to optimize (one response).
\end{compactitem}

\begin{rq2box}
Out of the 10 elicited prompt optimization patterns, the study participants admitted to mainly using those related to I/O and pre- and post-conditions. Improvements related to textual ambiguity were seldom adopted, and this was also the case for those aimed at driving code generation by providing I/O examples. 
\end{rq2box}

\subsection{RQ$_3$: Perceived usefulness of the guidelines}
Besides asking the study participants to report their usage of the prompt improvement patterns from our guidelines, we also asked them to evaluate their perceived usefulness. Results are reported in \figref{fig:usefulness}.

On the one side, the figure confirms, in terms of perceived usefulness, what the study participants have reported using. Indeed, explicitly specifying the \emph{I/O format}  is perceived as useful by almost all participants (88\%), and high percentages are also reported for \emph{pre-conditions} and \emph{post-conditions}. Similarly, some very specific improvement patterns were considered relatively less useful, consistent with their limited usage. These include, for example, a lack of coherent \emph{variable} mentioning, making the handling of \emph{exceptions} explicit, or better specifying \emph{unclear conditions}. For the last two cases, this confirms their applicability in a limited proportion of scenarios. 

On the other side, some guidelines whose patterns were reported to be rarely used were, instead, considered quite useful. Above all, this was the case of providing \emph{more examples}. This is a confirmation of the fact that their limited usage may simply depend on a lack of awareness.

We also observed a relatively high perception of usefulness, somewhat contrasting with the relatively lower usage, for better specifying dependency \emph{requirements}, as well as for clarifying \emph{ambiguities}. 

\begin{rq3box}
The study participants consistently perceived as useful some of the prompt improvement patterns they actually use, especially those related to pre- and post-conditions and I/O format. At the same time, they found useful guidelines related to driving the LLM by adding more I/O examples, but they also appreciated the need to avoid textual ambiguities.  
\end{rq3box}

\section{Threats To Validity}
Threats to \emph{construct validity} concern the relationship between theory and observation, and may arise both in the prompt elicitation phase and in their validation. Regarding the elicitation, we avoided considering cases in which the LLMs tried to overfit test cases as an improvement pattern. Regarding the validation, our study does not measure the guideline usage effectiveness or efficacy, nor does it provide exact measures of actual usage. Instead, we leverage self-reported use admission, which could be imprecise. Similarly, the perceived usefulness is not intended to represent an actual usefulness measured during a development task, and therefore can be affected by the participants' subjectivity.

Threats to \emph{internal validity} concern factors internal to our study that may affect our findings. As we explained in \secref{sec:methodology} during the automated code optimization, we used proper precautions to mitigate effects related to nondeterminism. We mitigated errors and subjectivity in our manual guideline elicitation by having two evaluators for each instance, who reviewed and resolved inconsistent cases.  Note that agreement by chance may not influence the conclusions of our study, because in RQ$_1$ we were merely interested in identifying possible improvement patterns, rather than in exactly assigning patterns to tasks. The patterns were then validated with the study reported in RQ$_2$ and RQ$_3$.

Regarding the guideline evaluation, we mitigated threats related to the questionnaire design by following proper survey design guidelines \cite{groves2009survey,pfleeger2001principles} and by having authors not involved in its design to validate it. Concerning the study participants, while their demographics indicate a good diversity in experience and expertise, we cannot exclude a self-selection bias. 

Threats to \emph{external validity} concern the generalizability of our findings. Concerning the prompt elicitation, two key elements may limit the generalizability of our findings. One is related to the choice of the three used benchmarks (BigCodeBench, HumanEval+, and MBPP+), and another to the specific programming language considered (Python). For the former, we have leveraged general-purpose and diverse benchmarks in terms of code tasks. However, we cannot exclude the possibility that leveraging other benchmarks may lead to further guidelines. Also, while our guidelines are totally language-agnostic, there could be improvement patterns that apply to some specific languages than others.
These threats were mitigated through the validation with practitioners, by asking them whether we missed some improvement patterns.

Last but not least, our study participants may not represent a wide population of developers, especially very expert ones. However, they cover a wide range of experience and expertise that may represent well enough practitioners interacting with LLMs for code generation.

\section{Implications}
This section discusses implications deriving from our work that can affect different kinds of stakeholders.

Concerning software development \textbf{practitioners}, the elicited guidelines can constitute a reference point to be followed when interacting with LLMs during code generation. While some optimizations (\eg those related to I/O, pre- and post-conditions, examples, or checking for textual ambiguity) may apply in most cases, developers may adopt other guidelines in specific circumstances. For example, those related to requirements may be applied once the technology to be used is known, whereas those pertaining to exceptions can be part of the inputs to apply to achieve better code robustness (\eg on the line of what related work did for code preferences \cite{weyssow2025}), but once again, when enough contextual details are known. In general, creating a decision tree explaining to developers which elements to improve in their prompts might be desirable, and when. 

What was said above also translates into implications for \textbf{educators}. Prompt engineering is nowadays becoming part of learning elements in curricula \cite{10578596,11024336}.  To this extent, our guidelines can represent a key component to be used to instruct software engineers in doing prompt engineering for code generation.

\textbf{Tool creators} and, in general, \textbf{researchers} working on creating novel recommenders for software development, could leverage our results to develop better tools. For example, tools could automatically analyze the prompt and the surrounding context, provide suggestions to the developers, and try to improve the prompt automatically if they realize that some needed elements are missing or unclear. 

\section{Conclusions and Future Work}
In this paper, we have empirically defined a catalog of prompt optimization guidelines for LLM code generation.
The guidelines have been defined by (i) iterating the interaction with the LLMs to make it pass tests and then asking it to optimize the prompt based on the successful iteration, and (ii) performing a manual analysis of the optimization differences between the initial and final prompts. We obtained 10 code-specific prompt optimization guidelines, covering different aspects,  such as I/O formatting, pre- and post-conditions, I/O examples, dependencies, implementation aspects (algorithms, exceptions, complex conditions), and different forms of ambiguities.

We have validated the guideline catalog through a study with 50 practitioners, investigating the extent to which they (i) already use the prompt optimization patterns in our guidelines, and (ii) perceive them to be useful. On the one hand, results of the study indicate that participants consider some of the elicited guidelines---such as those related to I/O formatting and pre-post conditions---as part of their prompt engineering activity, whereas those pertaining to ambiguities and I/O examples were less used. On the other hand, they perceived the latter particularly useful, along with those they already use.

The study leads towards implications for different stakeholders, primarily developers and educators, who can leverage the guidelines to improve their code or to instruct new practitioners about code-specific prompt engineering, but also researchers and tool creators, as they could leverage the guidelines to craft tools for automated code generation, prompt guidance, and optimization. 

Work-in-progress goes towards different avenues. We would like to extend our empirical investigation to other languages and tasks. Also, we plan to conduct controlled experiments to assess the actual guidelines' usefulness. Finally, as mentioned above, it would be worthwhile to leverage the guidelines to create tools for automated prompt quality assessment and improvement.

\section{Data Availability}
The study dataset is available in our replication package \cite{replication}.

\balance
\bibliographystyle{ACM-Reference-Format}
\bibliography{Biblio.bib}
\end{document}